\documentclass{aa} 
\usepackage{times,amssymb,amsmath} 
\usepackage{epsfig} 
\usepackage{lscape}

\newcommand{\mrm}{\mathrm}

\begin{document} 

\title{A multi-wavelength study of $z = 3.15$ Lyman-$\alpha$ emitters in the 
GOODS South Field
\thanks{Based on observations carried out at the European Southern
Observatory (ESO) under prog. ID No. 70.A-0447, 274.A-5029 and LP168.A-0485.}}

\author{K.K. Nilsson\inst{1,2}
        \and P. M\o ller\inst{1}
        \and O. M{\"o}ller\inst{3}
        \and J.~P.~U. Fynbo\inst{2}
        \and M.J. Micha{\l}owski\inst{2}
        \and D. Watson\inst{2}
        \and C. Ledoux\inst{4}
        \and P. Rosati\inst{1}
        \and K. Pedersen\inst{2}
        \and L.F. Grove\inst{2}
}

\institute{
   European Southern Observatory, Karl-Schwarzschild-Stra\ss e 2, 85748
   Garching bei M\"unchen, Germany\\
\and
   Dark Cosmology Centre, Niels Bohr Institute, University of Copenhagen, 
   Juliane Maries Vej 30, 2100 Copenhagen $\O$, Denmark\\
\and
   Max-Planck-Institut f{\"u}r Astrophysik, Karl-Schwarzschild-Stra\ss e 1, 
   85741 Garching bei M\"unchen, Germany\\
\and
   European Southern Observatory, Alonso de C\'ordova 3107, Casilla
   19001, Vitacura, Santiago 19, Chile\\
}
\offprints{kim@dark-cosmology.dk}
\date{Received date / Accepted date}
\titlerunning{Ly-$\alpha$ emitters in GOODS-S}

\abstract{
Ly$\alpha$-emitters have proven to be excellent probes of faint, star-forming
galaxies in the high redshift universe. However, although the sample of known 
emitters
is increasingly growing, their nature (e.g. stellar masses, ages, 
metallicities,
star-formation rates) is still poorly constrained.}
{
We aim to study the nature of Ly$\alpha$-emitters, to find the properties
of a typical Ly$\alpha$-emitting galaxy and to compare these properties
with the properties of other galaxies at similar redshift, in 
particular Lyman-break galaxies.}
{
We have performed narrow-band imaging at the VLT, focused on Ly$\alpha$ at 
redshift
$z \approx 3.15$, in the GOODS-S field. We have identified a sample of 
Ly$\alpha$-emitting candidates, and we have studied their Spectral Energy
Distributions (SEDs).}
{
We find that the emitters are best fit by an SED with low metallicity 
($Z/Z_{\odot} = 0.005$) , low 
dust extinction (A$_V \approx 0.32$) and medium stellar masses 
of approximately $10^9$~M$_{\odot}$. The age is not very well constrained. 
One object out of 24 appears to be a high redshift Ly$\alpha$-emitting dusty
starburst galaxy. We find filamentary
structure as traced by the Ly$\alpha$-emitters at the 
4$\sigma$ level. The rest-frame UV SED of these galaxies 
is very similar to that of Lyman Break Galaxies (LBGs) and comply 
with the selection criteria for $U$-band drop-outs, except
they are intrinsically fainter than the current limit for LBGs.}
{
Ly$\alpha$-emitters are excellent probes of galaxies in the distant universe,
and represent a class of star-forming, dust and AGN free, medium mass objects.}

\maketitle

\section{Introduction}
The possibility to use the Ly$\alpha$ emission line to study galaxies in early
stages of their formation was outlined already by Partridge \& Peebles (1967)
nearly 40 years ago, but early surveys (see Pritchet 1994 for a review) failed
to produce anything other than upper limits. The unexpected faintness of the
objects caused it to take almost three decades before the narrow-band technique
was successfully used to identify the first high redshift Ly$\alpha$ emitting 
galaxies that were not dominated by Active Galactic Nuclei (e.g.,
Lowenthal et al. 1991; M\o ller \& Warren 1993; Hu \& McMahon 1996; Petitjean et
al. 1996; Francis et al. 1996; Cowie \& Hu 1998). It is only recently, with 
the advent of 8 m
class telescopes and sensitive detectors, that larger samples of Ly$\alpha$
selected objects have been reported (e.g. Steidel et al. 2000; Malhotra \&
Rhoads 2002; Fynbo et al. 2003; Ouchi et al. 2003; Hayashino et al. 
2004; Venemans et al. 2005).

Already during the early studies, two interesting suggestions were raised.  
First, it
was found that there is a tendency for Ly$\alpha$ selected objects to ``line
up'' as strings in redshift space, and that they therefore may be excellent
tracers of filaments at high redshifts (Warren \& M\o ller 1996; Ouchi et al. 
2004; Matsuda et al. 2005). Secondly, they were found to
have very faint broad band magnitudes and therefore could be good 
tools in detecting faint, high redshift galaxies (Fynbo et al.
2001; Fujita et al. 2003; Venemans et al. 2005; Gawiser et al. 2006).

The first of those suggestions was explored theoretically via modelling of
structure formation including assignment of Ly$\alpha$ emission to the models
(Furlanetto et al. 2003; Monaco et al. 2005), and has been confirmed
observationally (M{\o}ller \& Fynbo 2001; Hayashino et al. 2004), who also 
proposed to use such
structures for a new cosmological test to measure $\Omega_{\Lambda}$ by looking
at the ``length-to-radius'' ratio of filaments observed from the side or
end-on. This proposed test was subsequently explored in detail by 
Weidinger et al. (2002).

The second suggestion has gained significant interest because there are now
several additional, but independent, ways of identifying high redshift but
optically faint galaxies, e.g. Damped Ly$\alpha$ Absorbers (DLA) galaxies 
(Wolfe et al. 1986; M\o ller et al. 2002; Wolfe, Gawiser \& Prochaska 2005), 
Gamma-Ray Burst (GRB) host
galaxies (Fruchter et al. 2006), sub-mm galaxies (Chapman et al. 2004). 
Most
galaxies from such searches are however too faint to be identified in current
ground-based optical flux limited samples. A significant project 
(the Building the Bridge Survey, BBS) aimed at
addressing those issues is currently underway at the ESO Very Large Telescope
(VLT) (Fynbo et al. 2001; Fynbo et al. 2003).  The very faintness of the
objects, however, renders it difficult to make any detailed comparisons. While
some DLA galaxies have been imaged with HST (Warren et al. 2001; M{\o}ller et
al. 2002) and likewise GRB hosts (Jaunsen et al. 2003; Fynbo
et al. 2005), and sub-mm galaxies (Smail et al. 2004; Pope et al. 2005; Schmitt
et al. 2006), only a very small subset of the Ly$\alpha$ selected galaxies have
been imaged with HST (Pascarelle et al. 1996; Venemans
et al. 2005; Overzier et al. 2006) and furthermore, such images are mostly too 
shallow for a detailed study.

The GOODS-S (Giavalisco et al. 2004) provides a unique opportunity to obtain a
deep, high resolution, multi-band data set of a complete and unbiased sample of
Ly$\alpha$-emitters (or LEGOs for Ly$\alpha$ Emitting Galaxy-building Objects;
M{\o}ller~\&~Fynbo (2001)).
We have therefore started a program to collect a complete, unbiased sample of
LEGOs in the GOODS-S. This allows a detailed study of the global properties, e.g.
photometry and morphology of LEGOs, as well as SED fits including photometry 
from the large, available multi-wavelength data-set.

This paper is organised as follows; in section 2 we present the
imaging observations, data reductions and candidate selection process. 
In section 3 we present spectroscopic observations of three candidates 
as well as the results from these observations. Sections 4, 5 and 6 contain the
discussion of various aspects of the LEGO candidate sample; first the basic 
characteristics of the LEGO sample, then the SED fitting and finally a 
comparison to Lyman-Break Galaxies. The conclusion is presented in section 7. 

\vskip 5mm
Throughout this paper, we assume a cosmology with $H_0=72$
km s$^{-1}$ Mpc$^{-1}$ (Freedman et al. 2001), $\Omega _{\rm m}=0.3$ and 
$\Omega _\Lambda=0.7$.
Magnitudes are given in the AB system. We survey a co-moving volume of
$\approx 3300$~Mpc$^3$.

\section{Imaging}
\subsection{Narrow band observations and data reduction}
\label{obs}
\begin{figure}[!ht] 
\begin{center} 
\epsfig{file=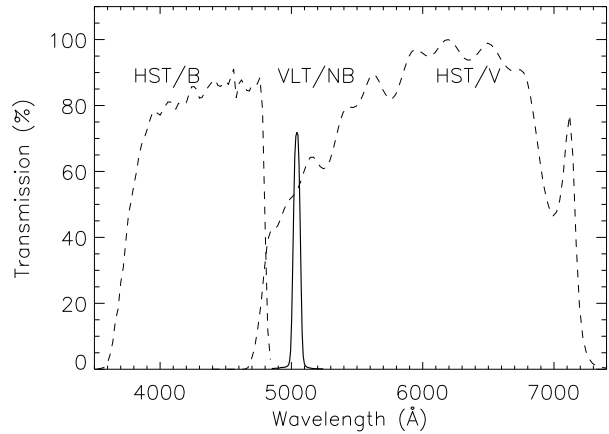,width=9.0cm}
\caption{Transmission of selection filters. The VLT FORS1 narrow-band 
filter is drawn with a solid line. Dashed lines show the HST $B$ and $V$ filters.}
\label{filters} 
\end{center} 
\end{figure}

A 400$\times$400 arcsec$^2$ section of the GOODS-S field centred on 
R.A.~$ = 03^h 32^m 21.9^s$  and Dec~$ = -27^{\circ} 45' 50$\farcs$7$ (J2000) 
was observed with FORS1 on the VLT 8.2 m telescope Antu
during two visitor mode nights on December 1-3, 2002. The log of
observations is given in Table~1.
A total of 16 dithered exposures were obtained over the two nights
for a combined exposure time of 30000 seconds, all with the narrow band
filter OIII/3000+51 and using the standard resolution collimator (0.2$\times$0.2
arcsec$^2$ pixels). For this setup the central wavelength of the
filter is 505.3 nm with a FWHM of 5.9 nm which corresponds to the
redshift range $z = 3.131-3.180$ for Ly$\alpha$.
The transmission curve of the filter is shown in Fig.~\ref{filters}.
The four spectrophotometric standards Feige110, LDS749B, LTT3864,
and LTT3218 were observed on the same nights.

\begin{figure*}[!ht] 
\begin{center}
\epsfig{file=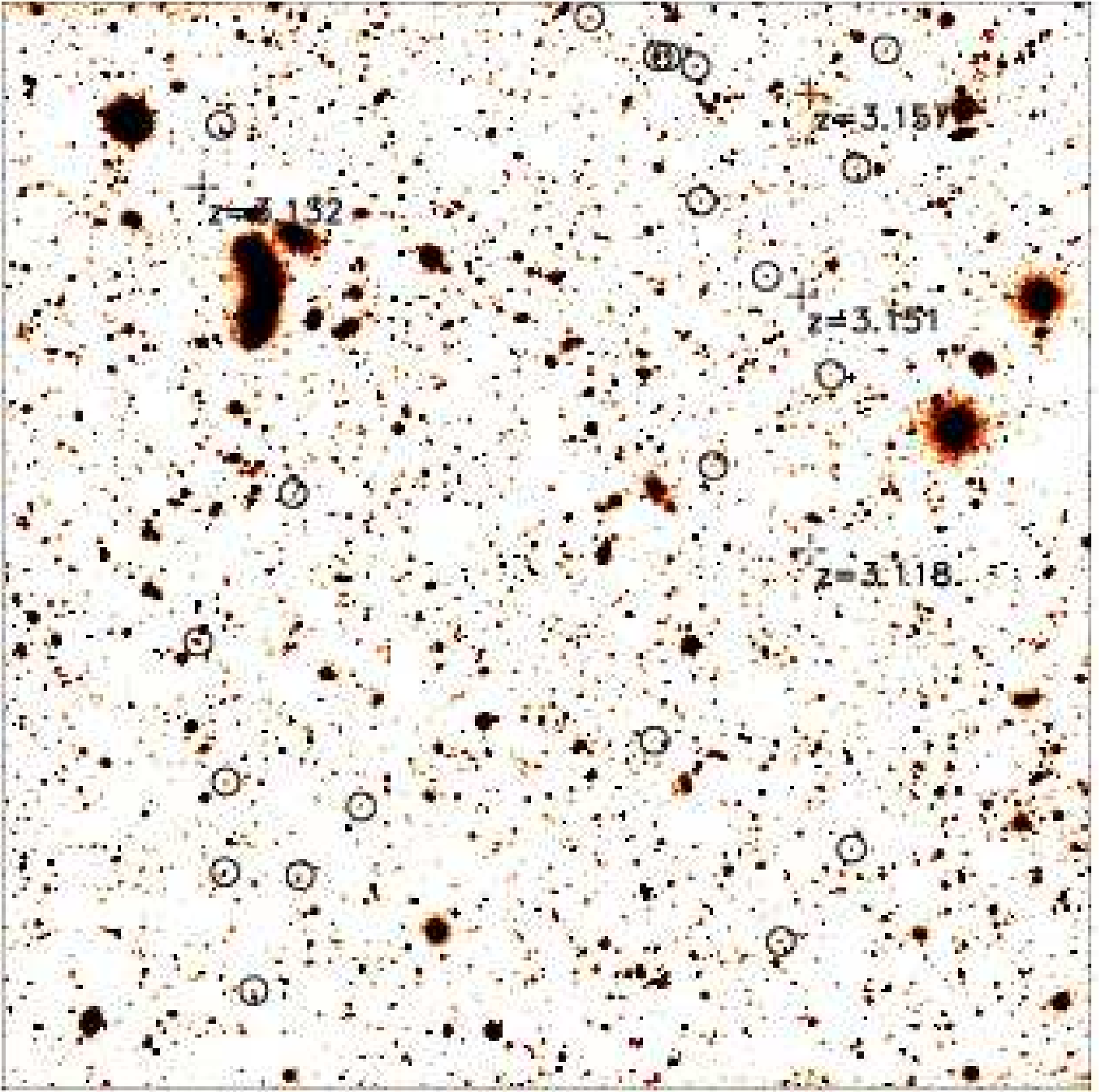,width=18.0cm,angle=90}
\caption{The VLT/narrow-band image of the 400$\times$400 arcsec$^2$ field with
the positions of selected LEGO candidates (see Sect.~\ref{LEGOs}) shown with 
circles. Spectroscopically confirmed candidates are marked with crosses and 
their redshifts are indicated. North is 
up and East is left. The right, uppermost spectroscopically confirmed candidate
is the Ly$\alpha$ blob (Nilsson et al. 2006).}
\label{field} 
\end{center} 
\end{figure*}

The observing conditions were unstable during the two nights with the
seeing FWHM, as measured on the images, varying between 0\farcs66 and 
1\farcs25 on the first night
and  between 1\farcs4 and 3\farcs3 on the second night.
The images were reduced (de-biased and corrected for CCD pixel-to-pixel
variations) using standard techniques. 
The individual reduced images were combined using a modified version
of our code that optimises the Signal-to-Noise (S/N) ratio for faint,
sky-dominated sources (see M\o ller \& Warren, 1993, for details on this
code). The 5~$\sigma$ detection limit of the combined narrow-band image 
as measured in circular apertures with radius twice the
full width half maximum of point sources, i.e. with radius 1\farcs6, is
mag(AB)$ = 26.1$. The combined narrow-band image is shown in Fig.~\ref{field}. 

\begin{table}[!ht] 
\begin{center}
\caption{Log of imaging observations with FORS1. }
\begin{tabular}{@{}lcccccc}
\hline
date & total exp. & seeing range & \\
\hline
01-02.12.2002   &   5.54 hours   &   0\farcs66-1\farcs25  &  \\
02-03.12.2002   &   2.78 hours   &   1\farcs43-3\farcs30    &  \\
\hline
\label{journal} 
\end{tabular} 
\end{center} 
\end{table}

\subsection{Selection of LEGOs in the fields}
\label{LEGOs}

For the selection of LEGO candidates we used the narrow-band image as detection
image and the HST/ACS $B$- (F435W) and $V$-band (F606W) images as
selection images. The HST data is part of the public data in GOODS-S 
(Giavalisco et al. 2004).  This selection set-up, with a broad-band filter on
either side of the narrow-band filter, appears to be one of the most efficient 
configurations for selection of emission-line objects (Hayes \& {\"O}stlin 
2006; Hayes priv. communication). The HST images were re-binned to the pixel 
size of the narrow-band image. Due to the smaller field-of-view of the HST/ACS 
images, the narrow-band image was cut into six sub-images to match the size of
the HST images.

Our selection method consists of three consecutive steps. First,
using the software package SExtractor (Bertin \& Arnouts 1996), we select 
all objects
identified in the narrow-band image. The narrow-band image is
scanned with a detection threshold equal to the background sky-noise
and requiring a minimum area of 5 connected pixels above this
threshold. Centred on each candidate object we then extract
photometry from the narrow, $V$, and $B$-band images using identical
circular apertures of $2''$ diameter.
Note that through this process we make no attempt to identify
broad band counterparts to the narrow band objects. The broad band
photometry is extracted in apertures defined solely from the centroid
positions in the narrow band image. There is always a small but finite
possibility that an unrelated for- or background object could fall
inside the $2''$ aperture which would complicate the search for emission
line objects. This complication is minimised by not re-centering the
aperture on objects in the broad band images.

The second step is to accept only candidates that are
detected at S/N$>5$ in the narrow band circular aperture and are found at
least 20 pixels from the edge of each image. This leaves us with a
catalogue of 2616 narrow band objects within the resulting 
$385 \times 400$~arcsec$^2$ field.  The term ``narrow band objects'' is used 
here only to
underline that while they are $5 \sigma$ detections in the narrow band
image, they may or may not have been detected in the broad band images.

\begin{figure}
\begin{center}
\epsfig{file=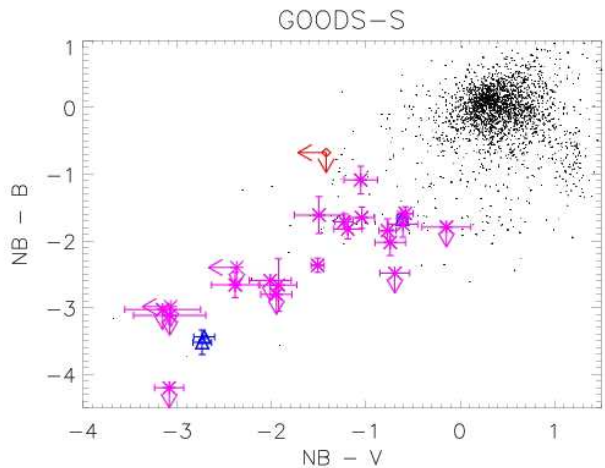,width=6.5cm,angle=90}
\caption{
Colour-colour plot. Dots mark the whole sample of 2616 objects selected with
SExtractor. Points with errors mark the selected emission-line objects. 
Stars (in magenta) mark 
the candidate sample, the triangles (in blue) the spectroscopically confirmed 
LEGOs and the open diamond (in red) the blob (Nilsson et al., 2006). Dots 
detected in the same region of the plot as
the selected sample, but that are not selected, consist of objects that were
discarded in the visual inspection.}
\label{colcol}
\end{center}
\end{figure}

In the third step we select the subset of our catalogue comprising of
potential emission line objects. Via interpolation of the flux levels
in the B and V bands it is easy to calculate the continuum flux at the
central wavelength of the narrow band filter, and from there to obtain the
equivalent width (EW) of a potential line and its associated propagated
statistical error. Note that we here
calculate the ``EW of the aperture'', which means that if there is only
one object in the aperture, then we find the EW of that object. If
there are additional unrelated neighbours inside the aperture, then
the calculated EW will be smaller than the actual EW. Our listed EWs
(see Table~\ref{selecttable})
are in that sense conservative lower limits.
For the present work we are interested only in those
with positive values of EW (emission line objects), and only the
subset of those where the significance of the line is high enough to
provide a high probability that it is reliable, i.e. providing a high
efficiency of spectroscopic follow-up work. For the current field we 
already have a few
spectroscopic confirmations (see Sec.~\ref{specsec}) and
we conservatively chose to cut at the EW significance level (2.9$\sigma$)
where all confirmed LEGOs are included. This corresponds to a
formal probability of 99.6\% for confirmation of each object and we 
find 106 such objects. Of these, three are associated with
a Ly$\alpha$-blob that we found in this field (Nilsson et al., 2006).
Ly$\alpha$-blobs (e.g. Fynbo et al. 1999; Steidel et al. 2000; 
Matsuda et al. 2004) are large luminous nebulae, with sizes up to 150 kpc,
emitting solely Ly$\alpha$ emission. The Ly$\alpha$ luminosity can reach
$10^{43}$~erg~s$^{-1}$.
79 of the 106 objects with EW excess are ruled out after visual inspection due to 
stellar artefacts, saturated 
sources or for lying on parts of the image where the HST sky background is 
poorly constrained.  
Thus, we are finally left with a sample of 24 LEGO candidates. These candidates 
are presented in Table~\ref{selecttable}. The formal probability of
99.6\% for confirmation of the emission line results in 0.10 spurious detections
in our sample of 24 objects. A colour-colour plot of the 
narrow-band sources can be found in Fig.~\ref{colcol}.

\begin{table}
\caption{Data on first selection. Coordinates are in J2000. Equivalent
widths are calculated from the ``first selection'', i.e. with 2$''$ radius
apertures, centred on the narrow-band source centroid, see section~\ref{LEGOs}.} S/N is the signal-to-noise of the equivalent width.
\label{selecttable}
\centering
\begin{tabular}{lcccccccc}
\hline\hline
LEGO & EW$_{\mathrm{obs}}$ & $\sigma_{\mathrm{EW}}$ & S/N  & R.A. & Dec\\
GOODS-S & (\AA) & (\AA) &  & &\\
\hline
1  & 896 & 57 & 16  & 03:32:14.83 & -27:44:17.5  \\
2  & 59  & 15 & 3.8 & 03:32:17.62 & -27:43:42.3 \\
3  & 172 & 24 & 7.1 & 03:32:18.56 & -27:42:48.4 \\
4  & 901 & 48 & 19  & 03:32:31.46 & -27:43:37.2 \\
5  & 184 & 30 & 6.0 & 03:32:30.02 & -27:48:37.1 \\
6  & 517 & 23 & 22  & 03:32:30.82 & -27:47:52.8 \\
7  & 153 & 32 & 4.7 & 03:32:13.40 & -27:47:43.9 \\
8  & 85  & 20 & 4.3 & 03:32:30.79 & -27:47:19.2 \\
9  & 100 & 24 & 4.2 & 03:32:12.49 & -27:42:45.8 \\
10 & 52  & 16 & 3.3 & 03:32:13.30 & -27:43:29.9 \\
11 & 154 & 33 & 4.6 & 03:32:27.03 & -27:47:28.3 \\
12 & 151 & 33 & 4.6 & 03:32:13.99 & -27:44:47.0 \\
13 & 53 & 18 & 2.9  & 03:32:14.58 & -27:45:52.5 \\
14 & 37  & 4  & 8.3 & 03:32:18.82 & -27:42:48.3 \\
15 & 88  & 11 & 7.9 & 03:32:31.56 & -27:46:26.9 \\
16 & 34 & 7.5 & 4.4 & 03:32:20.72 & -27:42:33.8 \\
17 & 202 & 46 & 4.4 & 03:32:28.93 & -27:45:31.5 \\
18 & 269 & 55 & 4.9 & 03:32:17.26 & -27:45:21.0 \\
19 & 47  & 12 & 3.9 & 03:32:15.80 & -27:44:10.3 \\
20 & 58 & 15 & 3.9  & 03:32:30.96 & -27:43:14.2 \\
21 & 53 & 16 & 3.4  & 03:32:28.73 & -27:47:54.1 \\
22 & 58 & 15 & 4.0  & 03:32:17.77 & -27:42:52.1 \\
23 & 362 & 57 & 6.3 & 03:32:15.37 & -27:48:18.5 \\
24 & 136 & 24 & 5.7 & 03:32:18.89 & -27:47:03.6 \\
\hline
\end{tabular}
\end{table}

\subsection{Continuum counterparts and final photometry}\label{finalphot}
\begin{figure*}
\begin{center}
\epsfig{file=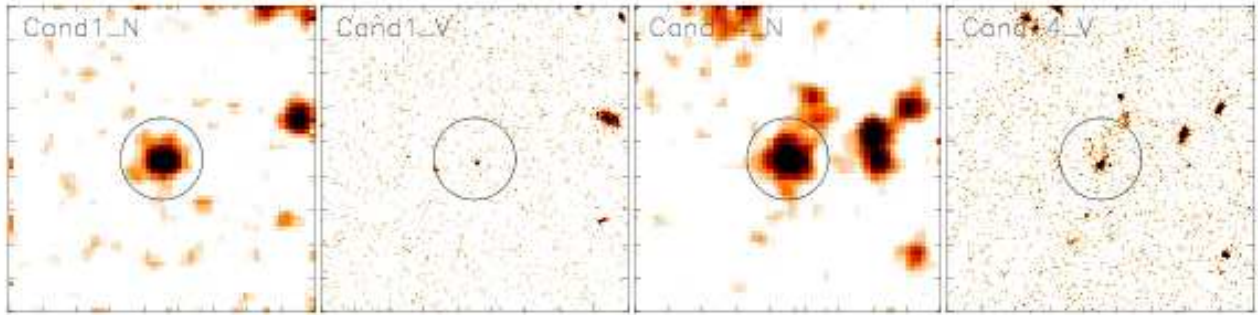,width=4.5cm,angle=90}
\caption{
In this figure, we illustrate the difficulty in determining which broad-band 
counterpart was associated with the narrow-band source. Images are $12''$ 
across and are centred on the narrow-band source. The circles mark the 
apertures used in the initial photometry, see section~\ref{LEGOs}. The two 
left panels show a simple case with
only one possible counterpart. The two right panels show a more complex case.
In this case, all objects within the circle were assumed to be counterparts.
In both cases a presumably unrelated object is seen at the edge of the 
aperture.}
\label{psfill}
\end{center}
\end{figure*}

Following the selection process outlined above, we examined the narrow-band and 
broad-band images to find continuum counterparts to the narrow-band objects.
The process of finding counterparts to the narrow-band objects is complicated 
by the very different PSF characteristics of the narrow- and broad-band images.
This is illustrated in Fig.~\ref{psfill}.
For some of the candidates, no obvious counterpart was detected, but rather 
several counterpart candidates were found with small offsets. To determine
what continuum objects were associated with the narrow-band source, two 
co-authors separately inspected the images visually. If only one 
counterpart was in the
vicinity of the centroid of the narrow-band source, this object was identified 
as
sole counterpart. If several sources were detected, their magnitudes were 
measured and the statistical probability that they would appear in the area
surrounding the narrow-band centroid was evaluated from number counts of 
galaxy searches. We then separately determined 
which counterparts we considered credible counterparts, and the lists were 
compared. Only counterparts assigned by both authors were accepted as 
counterparts. This yielded 2 LEGO candidates without counterparts, 14 
candidates with single counterparts and 8 LEGOs with two or more counterparts.
Aperture photometry was then performed on all 
candidates in the narrow--band and their selected counterparts in the 
HST broad band images. We used aperture radii of two times the 
FWHM of each image. Aperture corrections were calculated for each image for 
point sources (see Table~\ref{mwtab}). For candidates where multiple
components were assigned, small apertures were placed either \emph{i)}
centred on each counterpart separately if the counterparts are further apart
than two times the radius of the aperture, \emph{ii)} centred on a 
coordinate half way between the counterparts if the distance between 
counterparts is less than one times the radius of the aperture or
\emph{iii)} apertures slightly shifted from the central coordinate of the
counterparts to ensure that different apertures do not overlap if the distance
between counterparts is less than two, but more than one, times the aperture
radius. 
The magnitudes, EWs and star formation rates (SFRs) for all candidates can be 
found in Table~\ref{photometry}, see also Sec.~\ref{DiscLEGO}. 
Multiple candidates are marked with a star. To investigate how correct our
method is for measuring fluxes of multiple objects, we also measured the
photometry using larger apertures; for the six multiple 
counterparts with distances between the counterparts less than two times the
radius of the aperture (LEGO~GOODS-S\_\#~9, 10, 12, 14, 19 and 22), we 
applied new apertures with radii half of the distance between counterparts
plus the original aperture radius. For five candidate counterparts, the
difference in flux measured was within 1.5$\sigma$ of the previously measured
value. Hence we conclude that our original measurements are correct. The 
remaining candidate, LEGO~GOODS-S\_\#~14, has a very complicated morphology,
see also Fig.~\ref{psfill}.
For this candidate, the flux increased to M$_B = 25.49 \pm 0.11$, 
M$_V = 24.47 \pm 0.04$, M$_I = 24.36 \pm 0.13$ and M$_{z'} = 24.49 \pm 0.13$.
This reduces the observed EW to $91 \pm 8$~{\AA}. 

In summary we first performed simple circular aperture photometry in
order to select candidates based on our conservative emission-line
definition (Table~\ref{selecttable}). For the selected candidates, we 
then searched for
continuum counterparts in the high resolution HST images and carried
out detailed final photometry where such counterparts were found. This
final photometry provides the relevant magnitudes that describe the
objects and is reproduced in Table~\ref{photometry}.

\begin{table*}
\caption{Final photometry of candidates in narrow- and broad-bands 
(Sec.~\ref{finalphot}) and observed Ly$\alpha$ EWs. Magnitudes are 
calculated using two times full width half maximum apertures, including aperture
corrections, centred on each identified counterpart. Errors
are $1 \sigma$, upper limits are $3 \sigma$. Equivalent widths are
calculated again from the magnitudes printed in this table. Star formation 
rates are from the 
Ly$\alpha$ fluxes. Emitters marked in bold are spectroscopically confirmed, see
Sec.~\ref{specsec}. Candidates marked with a star have multiple counterparts.
For these candidates, the total magnitude is given here. }
\label{photometry}
\begin{tabular}{c|c|c|c|c|c|c|c}
\hline
\hline
LEGO & M$_{narrow}$ & M$_B$ & M$_V$ & M$_i$ & M$_{z'}$ & EW$_{\mathrm{obs,Ly}\alpha}$ ({\AA}) & SFR$_{\mathrm{Ly}{\alpha}}$ (M$_{\odot}$/yr) \\
GOODS-S{\_}\# & & & & & & & \\
\hline
\bf{1} &$\bf{24.35 \pm 0.07}$ & $\bf{27.87 \pm 0.17}$ & $\bf{27.08 \pm 0.07}$ & $\bf{27.33 \pm 0.28}$ & $\bf{27.07 \pm 0.22}$ & $\bf{1006 \pm 71}$ & $\bf{3.94 \pm 0.24}$  \\
2 & $25.36 \pm 0.17$ & $28.02 \pm 0.35$ & $27.29 \pm 0.10$ & $27.39 \pm 0.26$ & $ > 27.52$       & $434 \pm 82$ & $1.55 \pm 0.22$   \\
3 & $24.97 \pm 0.06$ & $ > 28.08      $ & $28.05 \pm 0.38$ & $ > 27.67      $ & $ > 27.24$       & $912 \pm 52$ & $2.22 \pm 0.11$  \\
\bf{4} & $\bf{24.19 \pm 0.02}$ & $\bf{27.63 \pm 0.02}$ & $\bf{26.91 \pm 0.11}$ & $\bf{27.25 \pm 0.30}$ & $\bf{27.07 \pm 0.32}$ & $\bf{956 \pm 15}$ & $\bf{4.55 \pm 0.07}$  \\
5 & $25.29 \pm 0.03$ & $ > 28.32      $ & $28.45 \pm 0.40$ & $ > 28.68      $ & $ > 26.89      $ & $881 \pm 27$ & $1.65 \pm 0.04$ \\
6 & $24.03 \pm 0.04$ & $ > 28.23    $   & $27.12 \pm 0.15$ & $ > 27.52      $ & $ > 27.31$       & $1639 \pm 64$ & $5.28 \pm 0.19$ \\
7 & $25.45 \pm 0.21$ & $ > 28.05    $   & $27.47 \pm 0.09$ & $27.34 \pm 0.28$ & $ > 27.26$       & $436 \pm 103$ & $1.42 \pm 0.24$ \\
8 & $25.23 \pm 0.14$ & $27.89 \pm 0.13$ & $27.62 \pm 0.21$ & $27.39 \pm 0.37$ & $ > 27.37$       & $532 \pm 81$ & $1.74 \pm 0.21$ \\
9$^{\star}$ & $24.75 \pm 0.08$ & $26.46 \pm 0.09$ & $25.98 \pm 0.06$ & $25.64 \pm 0.13$ & $26.03 \pm 0.15$ & $169 \pm 16$ & $2.73 \pm 0.18$ \\
10$^{\star}$ & $25.02 \pm 0.12$ & $26.84 \pm 0.09$ & $26.21 \pm 0.09$ & $26.24 \pm 0.08$ & $26.50 \pm 0.20$ & $179 \pm 27$ & $2.13 \pm 0.22$ \\
11 & $25.28 \pm 0.16$ & $ > 28.27     $  & $ > 28.36      $ & $ > 27.86      $ & $ > 27.60$       & $ > 834    $ & $1.66 \pm 0.22$ \\
12$^{\star}$ & $25.97 \pm 0.24$ & $ > 27.76     $ & $26.11 \pm 0.09$ & $27.43 \pm 0.29$ & $ > 27.46       $ & $76 \pm 34$ & $0.89 \pm 0.18$ \\
\bf{13} & $\bf{25.63 \pm 0.23}$ & $\bf{27.24 \pm 0.15}$ & $\bf{27.13 \pm 0.12}$ & $\bf{27.24 \pm 0.28}$ & $\bf{ > 27.08}$ & $\bf{178 \pm 57}$ & $\bf{1.21 \pm 0.23}$ \\
14$^{\star}$ & $23.97 \pm 0.05$ & $25.64 \pm 0.05$ & $24.58 \pm 0.03$ & $24.39 \pm 0.04$ & $24.64 \pm 0.06$ & $110 \pm 8$ & $5.58 \pm 0.26$ \\
15$^{\star}$ & $24.22 \pm 0.05$ & $26.59 \pm 0.09$ & $25.73 \pm 0.04$ & $25.25 \pm 0.04$ & $25.22 \pm 0.06$ & $296 \pm 18$ & $4.42 \pm 0.21$ \\
16 & $24.92 \pm 0.11$ & $ > 27.72      $ & $26.87 \pm 0.12$ & $26.12 \pm 0.08$ & $25.46 \pm 0.08$ & $473 \pm 59$ & $2.33 \pm 0.23$ \\
17$^{\star}$ & $25.09 \pm 0.08$ & $26.94 \pm 0.16$ & $25.86 \pm 0.05$ & $25.62 \pm 0.08$ & $25.33 \pm 0.08$ & $138 \pm 15$ & $1.98 \pm 0.14$ \\
18 & $25.37 \pm 0.16$ & $27.39 \pm 0.12$ & $26.11 \pm 0.04$ & $25.73 \pm 0.04$ & $25.84 \pm 0.10$ & $149 \pm 32$ & $1.54 \pm 0.21$ \\
19$^{\star}$ & $25.43 \pm 0.07$ & $27.02 \pm 0.06$ & $26.01 \pm 0.03$ & $25.80 \pm 0.08$ & $26.16 \pm 0.07$ & $102 \pm 11$ & $1.45 \pm 0.10$ \\
20 & $25.99 \pm 0.15$ & $27.08 \pm 0.14$ & $27.04 \pm 0.09$ & $26.23 \pm 0.11$ & $26.02 \pm 0.13$ & $90 \pm  23$ & $0.87 \pm 0.11$ \\
21 & $25.48 \pm 0.06$ & $ > 27.88      $ & $ > 27.86    $   & $ > 27.44      $ & $ > 27.35      $ & $ > 443    $ & $1.38 \pm 0.07$ \\
22$^{\star}$ & $25.05 \pm 0.15$ & $26.80 \pm 0.12$ & $25.65 \pm 0.04$  & $25.65 \pm 0.08$ & $25.60 \pm 0.08$ & $115 \pm 26$ & $2.07 \pm 0.27$ \\
23 & $25.30 \pm 0.15$ & $ > 27.78      $ & $25.99 \pm 0.05$ & $25.47 \pm 0.04$ & $25.43 \pm 0.07$ & $174 \pm 35$ & $1.64 \pm 0.21$ \\
24 & $25.46 \pm 0.12$ & $27.12 \pm 0.10$ & $26.51 \pm 0.06$ & $26.02 \pm 0.07$ & $26.04 \pm 0.12$ & $146 \pm 25$ & $1.41 \pm 0.15$ \\
\hline
\end{tabular}
\end{table*}

\section{Spectroscopy}\label{specsec}
\subsection{Observations and reductions}
Follow-up Multi-Object Spectroscopy (MOS) was obtained in service mode with 
FORS1/VLT UT2 over the time period December 2004 -- February 2005. The total 
observing time of 5.5 hours was granted to confirm the redshift of the 
Ly$\alpha$-blob found in this field, as published in Nilsson et al.~(2006).
In addition, we had the opportunity to add three of our compact emitters on 
the mask.
The mask preparation was done using the \emph{FORS Instrumental Mask Simulator}.
Stars were placed on the remaining slits for calibration purposes. The 
MOS slitlets had a width of 1\farcs4 and the combination of grism 600V and order 
sorting filter GG435 was used. The grism covers the wavelength range 
4650 {\AA} to 7100 {\AA} with a resolving power of approximately 700.
 The seeing varied between 0\farcs77 -- 1\farcs2.

The bias subtraction, flat-fielding and wavelength calibration was performed 
using the FORS1 pipeline. The individual, 2--dimensional, reduced science 
spectra were combined using a $\sigma$--clipping for rejection of cosmic ray 
hits. The sky was subsequently subtracted with MIDAS by averaging the values 
of all pixels on either side of the spectrum and expanding this value to the 
size of the frame. One--dimensional spectra were extracted by summing the 
column values over the spectra. The spectra were then flux calibrated with 
three stars that were on our slits, by measuring the stellar fluxes in the 
narrow-band image in $2''$ diameter apertures and comparing to the integrated 
flux in the narrow-band from the 1--d spectra. Because LEGOs are often extended
in Ly$\alpha$ (M{\o}ller \& Warren 1998; Fynbo et al. 2001) they are likely 
to have higher Ly$\alpha$ fluxes than those listed in Table~\ref{spectable}.

\subsection{Results of first spectroscopic follow-up}
\begin{figure*}[!ht] 
\begin{center}
\epsfig{file=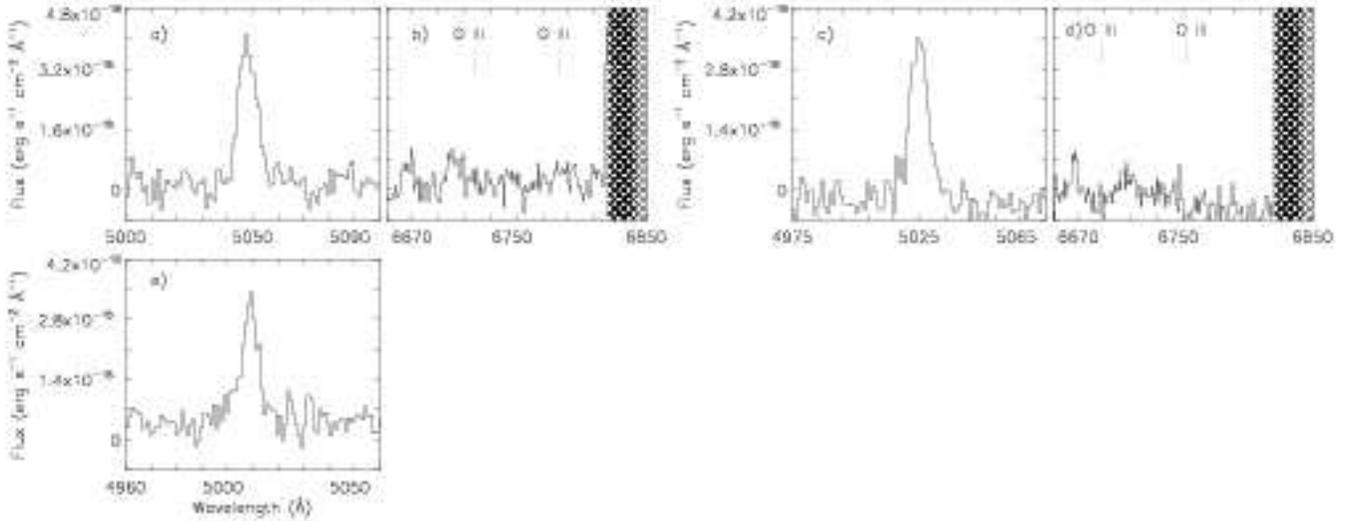,width=7.5cm,angle=90}
\caption{Spectra of confirmed LEGOs in GOODS-S. \emph{a)} LEGO\_GOODS-S\#1 
emission line, \emph{b)} LEGO\_GOODS-S\#1 expected position of [OIII] lines if 
the detected emission line is [OII], \emph{c)} LEGO\_GOODS-S\#4
emission line, \emph{d)} LEGO\_GOODS-S\#4 expected position of [OIII] lines if 
the detected emission line is [OII],\emph{e)} LEGO\_GOODS-S\#13
emission line. Our spectrum did not cover the position of [OIII] if the 
emission line is [OII] for LEGO\_GOODS-S\#13. Hatched areas
mark the positions of bright sky lines.}
\label{specandim} 
\end{center} 
\end{figure*}

The spectra of our three confirmed high-redshift Ly$\alpha$-emitters can be 
found in Fig.~\ref{specandim}. Neither candidate show any other emission lines 
in their spectra. To consider if the emission line could be [OII], we study 
line ratios compared to other lines that should be observed in such a case.
In Fynbo et al. (2001), the expected line 
ratios of H$\beta$, [OII], [OIII] and NeII for [OII]-emitters were presented. 
For the
spectroscopic sample, the 3$\sigma$ upper limits of the ratio 
log(F$_{[OIII]}$/F$_{[OII]}$) are given in Table~\ref{spectable} if the 
emission lines is [OII].
 The limit will be the same for all other lines, as none are detected. 
As in Fynbo et al. (2001), we conclude that these are highly unlikely values for 
[OII]-emitters, and that the detected emission line is redshifted Ly$\alpha$. 
Details from the spectroscopic follow-up is presented in Table~\ref{spectable}.
\begin{table}
\caption{Names, Ly$\alpha$-fluxes, redshifts and line ratios for 
spectroscopically confirmed LEGOs. The line ratios refer to the upper limit
to the [OIII] line, if the emission line is [OII].}
\label{spectable}
\centering
\begin{tabular}{cccccc}
\hline\hline
LEGO\_GOODS-S\# & F$_{\mathrm{Ly}\alpha}$ & $z$ & log(F$_{[OIII]}$/F$_{[OII]}$)\\
 & (erg s$^{-1}$ cm$^{-2}$) & &\\
\hline
1 & $3.31 \times 10^{-17}$ & 3.151 & $< -0.62$ \\
4 & $2.79 \times 10^{-17}$ & 3.132 & $< -0.54$ \\
13 & $2.84 \times 10^{-17}$ & 3.118 & $< -0.55$ \\
\hline
\end{tabular}
\end{table}

In the following sections we analyse the entire sample of confirmed LEGOs and 
LEGO candidates together. We expect the contamination of low redshift emitters
to be small. Our previous surveys have had spectroscopic success rates of 
75~-~90~\% (Fynbo et al. 2001; Fynbo et al. 2003). Hence, we expect that more
than 18 of our 24 candidates are true Ly$\alpha$-emitters.

\section{Basic characteristics of LEGOs}
\subsection{SFR, surface density and sizes}\label{DiscLEGO}
For our final sample of LEGO candidates, we calculate the star formation rate
(SFR) as derived from Kennicutt (1983) by:
\begin{equation}
\mathrm{SFR} = \frac{L_{\mathrm{H}\alpha}}{1.12 \times 10^{42}} \mathrm{M}_{\odot} \mathrm{yr}^{-1}
\end{equation}
where the H$\alpha$ luminosity, $L_{\mathrm{H}\alpha}$, is obtained with the 
conversion between Ly$\alpha$ and H$\alpha$ luminosities of Brocklehurst (1971) 
of L(Ly$\alpha$) = $8.7 \times$~L(H$\alpha$).
The SFR values of the LEGO candidates can be found in Table~\ref{photometry}. 
The mean SFR, as derived from the Ly$\alpha$-emission for all candidates is 
$1.8$~M$_{\odot}/\mrm{yr}$. The total SFR is 
$43$~M$_{\odot}/\mrm{yr}$, yielding a star formation rate 
density $\rho_{\mrm{SFR}}$ of $0.013$~M$_{\odot}/\mrm{yr}/\mrm{Mpc}^3$.
This value is in very good agreement with other results for high redshift
galaxies at this redshift of
e.g. Madau et al. (1996; 0.016~M$_{\odot}/\mrm{yr}/\mrm{Mpc}^3$), Steidel et 
al. (1999; 0.05~M$_{\odot}/\mrm{yr}/\mrm{Mpc}^3$) and Cowie \& Hu 
(1998; 0.01~M$_{\odot}/\mrm{yr}/\mrm{Mpc}^3$). The results from Steidel et al. 
(1999) has been obtained from integrating the extrapolated luminosity 
function down to a luminosity of $0.1$~$L_{\star}$, and the results are also
corrected for dust by multiplying by a factor of 4.7. Hence, their results 
uncorrected for dust is 
0.011~M$_{\odot}/\mrm{yr}/\mrm{Mpc}^3$. There is very good agreement
between the dust uncorrected measurements of the SFR density at $z \sim 3$. 

We find a surface density of LEGOs at redshift $z = 3.15, \delta z = 0.05$ in 
the GOODS-S field of 0.53 objects
per arcmin$^2$. We can compare this with all $V$ band sources in the 
GOODS-S field by extracting all sources with $V$ band magnitudes between 
25 -- 28 in the available online catalogue. We find $23885$ such sources in the 
entire GOODS-S area, covering approximately $160$~arcmin$^2$, corresponding to
a surface density of $149$~arcmin$^{-2}$. If we assume a homogeneous density of
LEGOs between redshift $3.0 - 3.5$, then the surface density of LEGOs, scaled
with our candidate sample, will be $\approx 5.3$~arcmin$^{-2}$. 
Thus, approximately 4~\% of all $V$-detected sources with a magnitude of 
$V = 25 - 28$ were selected as Ly$\alpha$-emitters in the redshift 
range $z = 3.0 - 3.5$. 

\begin{figure} 
\begin{center}
\epsfig{file=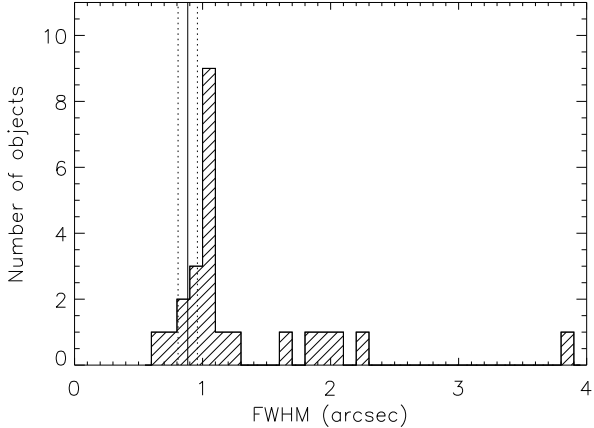,width=9.0cm,clip=}
\caption{Histogram plot over the size of our candidates in the narrow-band
image. The bin size is 0.1 arcsec, the solid line represents the PSF of the
image and the dotted lines the 1$\sigma$ error on the PSF. The PSF was determined
from 28 objects with SExtractor keyword CLASS\_STAR greater than 0.9 and fluxes
in the range of our LEGO candidates. The object in the 
highest FWHM bin is the GOODS-S blob (Nilsson et al., 2006).}
\label{extplot}
\end{center} 
\end{figure}

We measured the sizes of our candidates in the narrow-band images using the 
FLUX\_RADIUS option in
SExtractor. This gives the half width half maximum of each source. Object
LEGO\_GOODS-S\#~11 was excluded because it was blended with another, unrelated 
object. The histogram of the sizes, compared to a point source, is presented 
in Fig.~\ref{extplot}.
Most objects appear to be barely resolved. However, there is a 
tail of larger objects extending towards the GOODS-S blob.
In a future paper, we will present a complete morphological study of the 
candidate sample.

\subsection{Filamentary structure}
Previous studies of Ly$\alpha$ emitters have reported the identification
of filamentary structures (e.g. M{\o}ller \& Fynbo 2001; Hayashino et al. 2004;
Matsuda et al. 2005). The volume we survey here
is approximately 3 times as long along the line-of-sight as it is wide which 
means that several 
filaments could be crossing the volume at different redshifts and with
different position angles. If that was the case, they would likely
blend together to wash out individual structures. Bearing this in
mind we can still ask the question of whether the objects on our candidate
list are randomly distributed across the field, or if they appear to
be systematically aligned.

\begin{figure*} \begin{center}
\epsfig{file=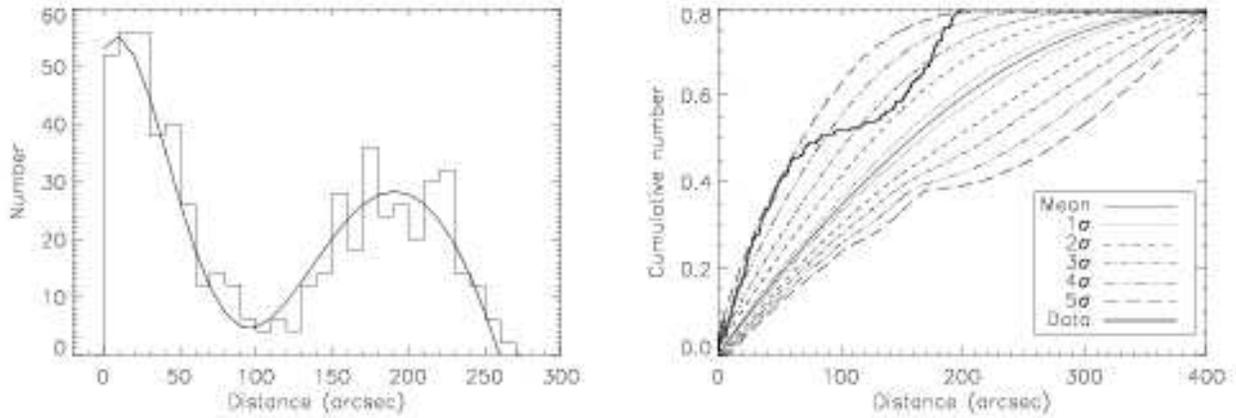,width=6.0cm,clip=,angle=90}
\caption{\emph{Left} Histogram distribution of distances between candidate 
objects in the x-direction, after re-alignment by $2.1^{\circ}$. Histogram is
binned over 50 pixels. Solid curve is a double Gaussian fit to the data. \emph{Right} Kolmogorov-Smirnov test of the distribution. Lines mark the simulated 
mean, and the 1 -- 5$\sigma$ contours. Thick line represents our data.}
\label{fil}
\end{center} \end{figure*}

Inspecting Fig.~\ref{field} we note that the candidates do in fact appear to be 
aligned in two filamentary structures approximately
along the y-axis of the CCD (oriented N-S). To investigate whether this is a 
significant 
effect, we have calculated the distances between all objects and plotted a 
histogram of the projected distances between all candidates in Fig.~\ref{fil}. 
The histogram shows a definite division in two components, one 
describing the typical width of a filament, and the other the projected 
distance between the filaments. 
To optimise the search, we calculated the angle of rotation that minimised the 
variance of the x-coordinates around a mean. For the two filaments, these angles
are $1.75^{\circ}$ and $2.52^{\circ}$ respectively. For the analysis, we rotate
the filament by the average, $2.13^{\circ}$. We then generated $10^7$
uniform random sets of coordinates, with the same number of objects, 
and repeated the same analysis of calculating distances between pairs.
To establish how reliable our observed distribution is, we perform a 
Kolmogorov-Smirnov test (e.g. Peacock 1983) on the
simulated distributions. The test showed that the likelihood of the alignment
being random is less than $2.5 \times 10^{-4}$, hence a near 4 sigma detection, 
see Fig.~\ref{fil}. We fitted a double Gaussian function to the histogram plot. 
This 
gives the typical width of the filaments as the FWHM of the first peak, and the 
distance between them as the mean distance to the second peak. We find that the
typical width is $\approx 250$~pixels, corresponding to 370 kpc at redshift 
3.15, in good agreement with the findings of M{\o}ller \& Fynbo (2001) who find 
a spectroscopically confirmed filament with 400 kpc radius from a set of
Ly$\alpha$-emitters at $z = 3.04$. The distance between 
the first and second peak is $\approx 950$~pixels, corresponding to 1.4 Mpc at 
this redshift. To further verify this filamentary structure would require more 
spectroscopic data, that would also enable a 3D-plot of the filaments in space.   

\section{SED fitting}
The imaging available in the \emph{GOODS-S} field is extensive. The 
data, in 14 publicly available broad-bands, used here is presented in 
Table~\ref{mwtab}.
\begin{table*}
\begin{center}
\caption{Deep, multi-wavelength data available in the GOODS-S field.  
The fifth column refers to the 3$\sigma$ detection limit in 
the sky in a $2 \times$~FWHM diameter aperture. The last column gives the 
3$\sigma$ detection limit as measured in a $2''$ radius aperture. }
\begin{tabular}{@{}lcccccccc}
\hline\hline
Filter/Channel & $\lambda_c$ & FWHM & Aperture radius & Aperture & 3$\sigma$ limit ($2 \times$~FWHM aperture) & 3$\sigma$ limit ($2''$ aperture)\\
& & & (arcsec) & correction & ($\mathrm{erg} \cdot \mathrm{cm}^{-2} \cdot \mathrm{s}^{-1} \cdot \mathrm{Hz}^{-1}$) & ($\mathrm{erg} \cdot \mathrm{cm}^{-2} \cdot \mathrm{s}^{-1} \cdot \mathrm{Hz}^{-1}$)\\
\hline
X-rays (\emph{Chandra})     &  4.15 keV    &  3.85 keV    & 2.25 & --- & $9.90 \cdot 10^{-34}$& $9.90 \cdot 10^{-34}$ &  \\
$U$ (\emph{ESO 2.2-m})        &  3630 \AA    &  760 \AA     & 3.00 & --- & $1.10 \cdot 10^{-30}$& $8.62 \cdot 10^{-31}$   & \\
$B$ (F435W, \emph{HST})              &  4297 \AA    &  1038 \AA    & 0.12 & 1.20 & $7.09 \cdot 10^{-32}$& $9.25 \cdot 10^{-30}$  &\\
$V$ (F606W, \emph{HST})              &  5907 \AA    &  2342 \AA    & 0.12 & 1.18 & $4.02 \cdot 10^{-32}$& $4.66 \cdot 10^{-30}$  & \\
\emph{i} (F814W, \emph{HST})              &  7764 \AA    &  1528 \AA    & 0.12 & 1.25 & $1.25 \cdot 10^{-31}$& $1.50 \cdot 10^{-29}$  & \\
$z'$ (F850LP, \emph{HST})              &  9445 \AA    &  1230 \AA    & 0.12 &1.34 & $1.88 \cdot 10^{-31}$& $3.00 \cdot 10^{-29}$  &\\
$J$ (\emph{VLT})              &  1.25 $\mu$m &  0.6 $\mu$m  & 0.60 & 1.22 & $1.78 \cdot 10^{-30}$& $5.31 \cdot 10^{-30}$ &\\
$H$ (\emph{VLT})              &  1.65 $\mu$m &  0.6 $\mu$m  & 0.60 & 1.21 & $4.11 \cdot 10^{-30}$& $1.86 \cdot 10^{-29}$  &\\
$Ks$ (\emph{VLT})             &  2.16 $\mu$m &  0.6 $\mu$m  & 0.60 & 1.37 & $4.06 \cdot 10^{-30}$& $1.56 \cdot 10^{-29}$  &\\
$Ch1$ (\emph{Spitzer})   &  3.58 $\mu$m &  0.75 $\mu$m & 1.30 & 2.41 & $1.14 \cdot 10^{-31}$ & $2.36 \cdot 10^{-30}$ & \\
$Ch2$ (\emph{Spitzer})   &  4.50 $\mu$m &  1.02 $\mu$m & 1.80 & 1.98 & $2.07 \cdot 10^{-32}$& $2.07 \cdot 10^{-30}$ &\\
$Ch3$ (\emph{Spitzer})   &  5.80 $\mu$m &  1.43 $\mu$m & 1.80 & 1.60 & $7.50 \cdot 10^{-29}$& $7.50 \cdot 10^{-30}$ &\\
$Ch4$ (\emph{Spitzer})   &  8.00 $\mu$m &  2.91 $\mu$m & 2.10 & 1.85 & $6.87 \cdot 10^{-30}$& $6.87 \cdot 10^{-30}$ &\\
$MIPS$ (\emph{Spitzer})  &  24.0 $\mu$m &  4.70 $\mu$m & 6.00 & --- & $1.23 \cdot 10^{-28}$ & $2.12 \cdot 10^{-29}$  \\
\hline
\label{mwtab}
\end{tabular}
\end{center}
\end{table*}
With this data-set, we wish to perform an SED fitting, in order to constrain 
properties such as stellar mass M$_\ast$, dust content A$_V$,
metallicity and age of the LEGOs. Only one of our 
candidates (LEGO\_GOODS-S\#16) is detected in bands other than the HST bands. This 
object is especially interesting as its SED is extremely red. It is excluded
from the SED fitting, and is discussed in Section~\ref{number19}. For the 
rest of the sample, the LEGOs are only detected in the HST bands and hence
we choose to stack the entire sample of 23 candidates. We can then draw 
conclusions on the general properties of this type of object. After stacking,
we get a faint detection in the $K_s$ band. The stacked magnitudes are 
given in Table~\ref{tabsed}. The lack of X-ray, MIPS 24$\mu$m and radio 
detections (no counterparts to any of our candidates to a 3$\sigma$ limit
of 24~$\mu$Jy, Kellermann et al. in preparation) 
indicates that the AGN fraction among these objects is low.
  
\begin{table}
\caption{Stacked magnitudes for the GOODS-S Ly$\alpha$-emitters. Errors
in the HST magnitudes were set to a conservative value of 0.08. Upper limits
are 3$\sigma$.}
\begin{tabular}{@{}lccc}
\hline
\hline
Band & Centr. Wavelength (\AA) & Obs. Magnitude & Mag. Error \\\hline
U & 3710 & $> 25.95$ & --- \\
B & 4297 & 27.57 & 0.08\\
N & 5055 & 24.96 & 0.08\\
V & 5907 & 26.74 & 0.08\\
\emph{i} & 7764 & 26.52 & 0.08\\
\emph{z} & 9445 & 26.56 & 0.08\\
J & 12500 & $> 26.28$ & --- \\
H & 16500 & $> 25.55$ & --- \\
Ks & 21500 & 25.26 & 0.29\\
Ch1 & 35800 & $> 23.06$ & --- \\
Ch2 & 45200 & $> 23.62$ & --- \\
Ch3 & 57200 & $> 24.37$ & --- \\
Ch4 & 79000 & $> 23.84$ & --- \\
\hline
\end{tabular}
\label{tabsed}
\end{table}

\subsection{Fitting method}
We used the GALAXEV code (Bruzual \& Charlot, 2003) to simulate composite
stellar populations, in order to fit the stacked SED of the LEGOs. The fitting 
was performed according to a Monte Carlo Markov Chain method (see e.g. Gilks et
al. 1995 for an introduction). 
In outline, the method works as follows; an initial set of parameter values is
chosen according to a uniform, random and logarithmic distribution within the allowed 
parameter space.
\begin{table}
\begin{center}
\caption{Parameter space sampled during the SED fitting. Metallicity is allowed
to have three different values ($Z/Z_{\odot} = 0.005$,~$0.2$~or~$1.0$). The
dust components are the two components of the Charlot \& Fall (2000) dust 
model used by GALAXEV. 
We fit the SED with a constant star forming model, where the star formation
rate in solar masses per year is given by ``SF-rate''.
}
\begin{tabular}{@{}lcccccccccc}
\hline
Parameter & Min. value & Max. value \\\hline
Metallicity ($Z/Z_{\odot}$) & 0.005 & 1.0 \\
Dust-$\tau$ & 0 & 4\\
Dust-$\mu$ & 0 & 1\\
SF-rate & 0.01 & 100\\
Age (Gyrs) & 0.001 & 1.5\\
\hline
\end{tabular}
\label{tabsedpars}
\end{center}
\end{table}
A summary of the parameter space is given in Table\,\ref{tabsedpars}. Given
the set of parameters, a corresponding $\chi^2$ value is calculated by
running the GALAXEV code, creating a
high-resolution spectrum with 6900 wavelength points from 91~{\AA}~to
160~$\mu$m. To obtain the magnitudes in each band, 
we apply the transmission curves for the filters of
the various observed wavebands; $U$, $B$, $V$, $i$, $z'$, $J$, $H$, $K_s$ 
and the four Spitzer bands, $Ch1 - Ch4$, 
In this analysis, we exclude the Spitzer $MIPS$ band as it is very 
difficult to stack images in this band because of the source confusion. At this
redshift, the $MIPS$ band is also contaminated by a PAH emission feature. 
We then
normalise the output spectrum so that the magnitude in the model $z'$-band equals
that of the observed $z'$-band. The $\chi^2$ is then calculated
by comparing the magnitudes in all the other bands.
We incorporate points with upper limits in the following way;
if the predicted flux lies below the upper limit, no value is added to the
total $\chi^2$, if the predicted flux lies above the
observed one, a $\chi^2$ is added, assuming the error on the upper limit is 0.1 in 
magnitude. Hence, models with flux above the upper limit may be acceptable if the
flux in this band is only slightly above the limit. In most cases though, the
model will be rejected due to high $\chi^2$.

Once the $\chi^2$ has been calculated for a particular model, a new
random set of parameters is chosen, by adding a ``step vector''. 
The step size in each parameter is chosen
randomly in a logarithmic interval between 1\% and 100\% of the total size of 
the parameter space. This step vector is equally likely to be positive or negative.
The choice of logarithmic step sizes is a natural choice if no assumptions 
about the scale
of change that affect the solution are to be made and ensures a fast
convergence. When a step has been calculated and the new parameters have been calculated, the new model
is accepted into the chain with a probability 
proportional to the exponential difference between the old and new $\chi^2$.
If the new
model is accepted, it is added as the next step in the chain, i.e. the parameters
of that particular model are printed in the output file. If the model is not
accepted,
a copy of the old parameters is added as the next step. This procedure is
repeated until a chain with 30000 elements has been created. The
independence of individual steps in the walk ensures
that the resulting chain is Markovian in character, i.e. that the resulting
chain after many iterations is a representation of the full probability
distribution function in the chosen parameter space. The output
file can then be used to study the distribution in each parameter, and to
determine the mean and the confidence levels within each parameter. It can also 
be used to study dependencies between parameters, such as e.g. in Fig.~\ref{contours}.

\subsection{Results from SED fitting}
\begin{figure}[!ht] 
\begin{center}
\epsfig{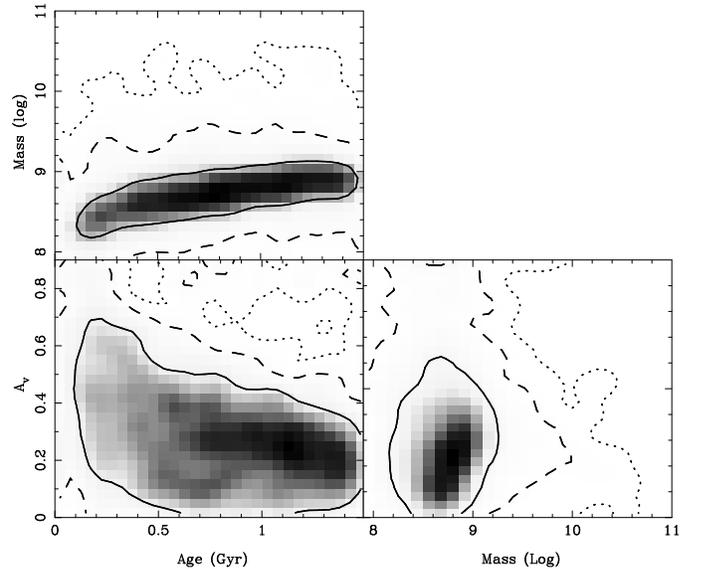}
\caption{Contour-plots of the mass, age 
and dust parameters in our SED fitting. Contours indicate 1, 2 and 3$\sigma$ 
levels.  }
\label{contours}
\end{center} 
\end{figure}

\begin{figure}[!ht] 
\begin{center}
\epsfig{file=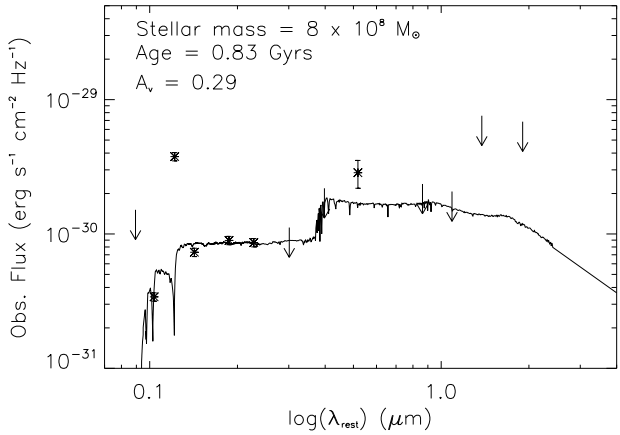,width=9.0cm,clip=}
\caption{The average spectrum of 10 models with $\chi^2\sim1$. The spectrum is
calculated by 1) creating the 10 spectra with the particular parameters of the
models with GALAXEV, 2) averaging the 10 spectra together, weighted with the $\chi^2$, 
so that the total flux density is $F_{\nu}=\left[ \sum_i F^i_{\nu}\times (1 / \chi^2_i) \right]/\sum_i (1/\chi^2_i)$. 
The parameters of the model spectrum are as indicated on the plot. They are 
the weighted average parameters of the 10 models used, weighted in the same way as the 
spectra. Data points from stacked SED. 
Upper limits are represented with arrows. The point well off the SED is the 
narrow-band magnitude. }
\label{plotsed}
\end{center} 
\end{figure}

The parameters we wish to fit are stellar mass, dust content,
star formation rate, metallicity and age. 
Redshift is set to 
be the central Ly$\alpha$ redshift of the narrow-band filter, i.e. $z = 3.15$.
We use the Salpeter initial mass function (IMF) from 0.1 to 100~M$_{\odot}$ 
and the extinction law of Charlot \& Fall (2000). We also 
incorporate the effect of the Ly$\alpha$ forest according to the model of
Madau (1995). We use constant star formation histories.
 The metallicity was allowed
to be either $Z/Z_{\odot} = 0.005$,~$0.2$~or~$1.0$. Of the 30000 runs, 73\% 
were 
with the lowest metallicity, 21\% with the medium metallicity and 6\% had solar 
metallicity. Hence the best fit models have a very low metallicity. 
For the rest of the analysis, we choose to only look at the models which
have the lowest metallicity, as these
models seem to be preferred. For these models, the best fit parameters are
M$_\ast = 4.7^{+4.2}_{-3.2} \times 10^8$~M$_{\odot}$, 
age~$ = 0.85^{+0.13}_{-0.42}$~Gyrs, 
A$_V = 0.26^{+0.11}_{-0.17}$, and star formation rate 
SFR~$ = 0.66^{+0.51}_{-0.31}$~M$_{\odot}$~yr$^{-1}$, where the errors are $1 \sigma$. Contour-plots
of the three parameters mass, age and dust are shown in Fig.~\ref{contours}.
The degeneracies between the different parameters can be seen. In 
Fig.~\ref{plotsed}, the weighted GALAXEV spectrum of a subset of 10 models 
with $\chi^2 \sim 1$ is shown with the measured SED 
overplotted. The $U$-band and Spitzer/IRAC data are too shallow to be useful 
in constraining the data, and the narrow-band data-point is plotted only for
reference.

\subsection{Object LEGO\_GOODS-S\#16}\label{number19}
\begin{figure*}[!ht] 
\begin{center}
\epsfig{file=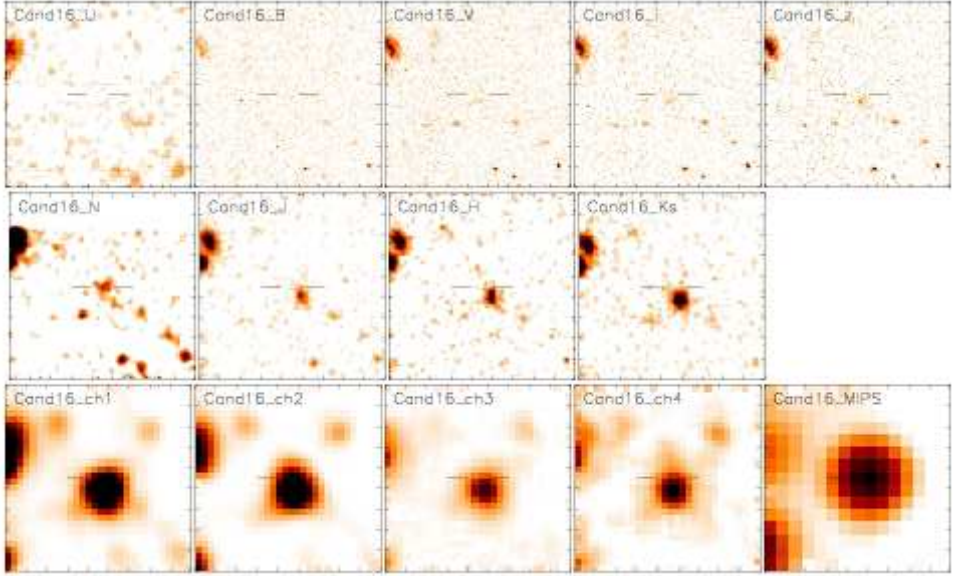,width=8.0cm,clip=,angle=90}
\caption{Thumb-nail images, $18''$ across, of LEGO\_GOODS-S\#16.}
\label{thumbs19}
\end{center} 
\end{figure*}

One candidate emission line object, LEGO\_GOODS-S\#16, was detected in all
available GOODS-S bands, except the $U$ and $B$ bands, and in X-rays. The fluxes of
the object can be found in Table~\ref{sed19}, and the thumb-nail images seen in
Fig.~\ref{thumbs19}.
As there is significant excess emission in our narrow-band filter, we expect 
the source to be 
either an [OII]-emitter at $z = 0.36$ or a Ly$\alpha$-emitter at $z = 3.15$.
\begin{table}
\caption{SED of object LEGO\_GOODS-S\#16. Upper limit is 3$\sigma$. 
Continuum sources are offset from the narrow-band source by approximately 
$0.8$~arcseconds, corresponding to 5.9 kpc at $z = 3.15$.}
\begin{tabular}{@{}lccc}
\hline
\hline
Band & Centr. Wavelength ($\mu$m) & Obs. flux ($\mu$Jy) & Flux Error ($\mu$Jy) \\\hline
B         &  0.430  &$> 0.03$  &   ---    \\
NB        &  0.506  &    0.39  &   0.042 \\
V         &  0.591  &    0.06  &   0.008 \\
\emph{i}  &  0.776  &    0.13  &   0.010 \\
\emph{z}  &  0.945  &    0.24  &   0.018 \\
J         &  1.25   &    3.47  &   0.157 \\
H         &  1.65   &    7.40  &   0.420 \\
Ks        &  2.15   &   12.51  &   0.565 \\
Ch1       &  3.58   &   24.87  &   0.188 \\
Ch2       &  4.50   &   30.86  &   0.200 \\
Ch3       &  5.80   &   31.85  &   0.710 \\
Ch4       &  8.00   &   21.31  &   0.990 \\
MIPS      & 24.00   &  271.11  &   7.001 \\
\hline
\end{tabular}
\label{sed19}
\end{table}
This object was first fit with the same type of stellar SED as the other 
sample, but 
with the redshift set to be either $z = 0.36$ if the narrow-band emission
is [OII] or $z = 3.15$ if the emission is Ly$\alpha$. This fitting yielded
no good fit, with $\chi^2 \gtrsim 500$.

\begin{figure}[!ht] 
\begin{center}
\epsfig{file=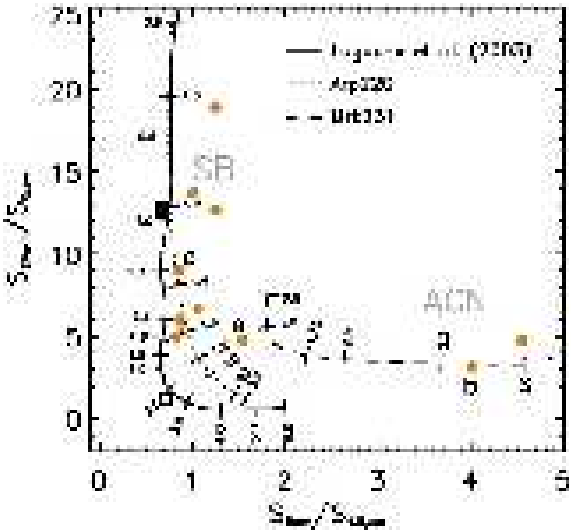,width=7.0cm,clip=,angle=90}
\caption{Diagram of Ivison et al. (2004). The x- and y-axes show the colours
in the Spitzer bands and the solid lines mark the locations of AGN and starburst
galaxies (SB) depending on redshift. The redshifts are marked along the lines. 
The orange dots mark the location of a set of sub-mm galaxies presented in Ivison et al. 
(2004). The solid square marks the location of the colours of LEGO\_GOODS-S\#16,
indicating a lower redshift starburst galaxy. The open square marks the colours
of this galaxy  if the MIPS 24~$\mu$m flux is decreased by a factor of 10 (see text).
This point is indicative of a redshift $z \sim 3$ starburst galaxy. }
\label{ivison}
\end{center} 
\end{figure}

\begin{figure*}[!ht] 
\begin{center}
\epsfig{file=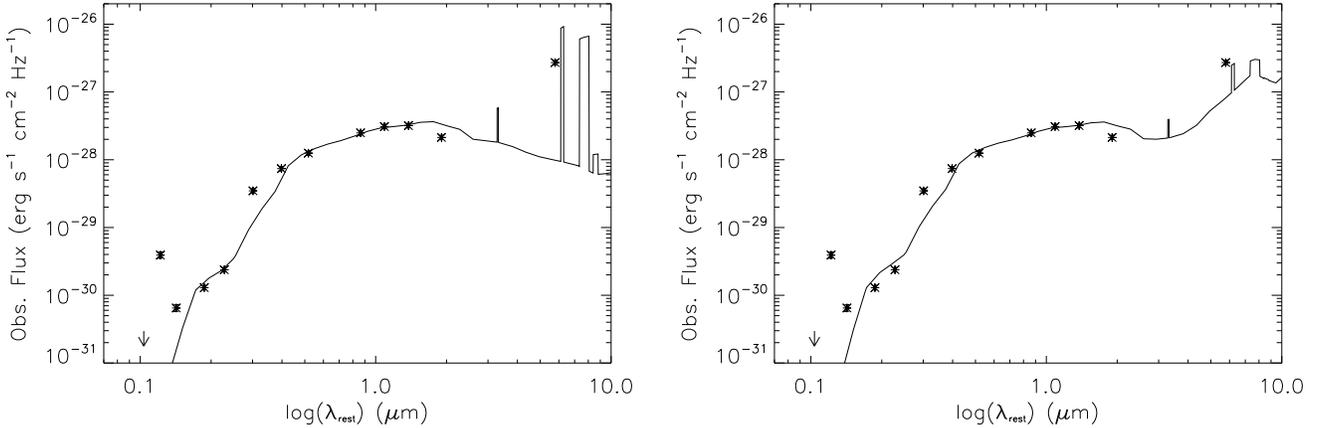,width=18.0cm,clip=}
\caption{Two preliminary GRASIL model fits to the SED of LEGO\_GOODS-S\#16, 
both with $\chi^2 \sim 50$. Points with error bars (of the same size as the 
point) are our data, with errors 
being purely statistical. The point off the curve at lower wavelengths is the
narrow-band detection.}
\label{grasilmodels}
\end{center} 
\end{figure*}

Two types of objects could show MIR colours similar to those of this galaxy; 
\emph{i)}
obscured AGNs (e.g. Lacy et al. 2004; Stern et al. 2005) and \emph{ii)} 
ULIRG/dusty starburst galaxies (e.g. Ivison et al. 2000; Klaas et al. 2001). 
The infrared colours can be plotted in the diagnostic colour-colour 
diagram of 
Ivison et al.~(2004), see Fig.~\ref{ivison}. 
In this diagram, Ivison et al.~(2004) plot 
the colours of Arp 220, Mrk 231 and a theoretical starburst spectrum as 
observed at different redshifts.
The comparison with the colours of LEGO\_GOODS-S\#16 shows the object 
to be more likely a
low-redshift starburst galaxy. However, Ivison et al. (2004) do not take PAH 
emission
into account. The most important PAH lines are at 3.3, 6.2, 7.7, 8.7, and 11.2 $\mu$m 
(corresponding to 13.7, 25.7, 32.0, 36.1, 46.5 at $z = 3.15$). Especially the second 
line falls on top of the MIPS 24~$\mu$m band. This would explain the extreme rise in 
flux in this band. In Fig.~\ref{grasilmodels}, we see that this emission could easily 
add a factor of ten to the flux in this band, hence the 24~$\mu$m/8~$\mu$m colour of 
LEGO\_GOODS-S\#16 can be adjusted downwards by a factor of ten. 
The extra line emission explains why a $z \sim 3$ star-burst galaxy would appear to be at 
lower redshifts.
To understand if this is a low or high redshift object, we have also attempted to get a 
photometric redshift estimate for this galaxy using the \emph{HyperZ} code
(Bolzonella et al. 2000), both with and without including the Spitzer data
points. The results were similar for both cases. When we do not include the 
Spitzer data,
the best fit is redshift $z = 1.7$ with a $\chi^2$ of approximately 8. For the 
redshift $z = 0.4$, the fit then has a $\chi^2 \approx 126$ and for the 
Ly$\alpha$ redshift of $z = 3.15$, the $\chi^2 \approx  35$. When the Spitzer 
points are included, the higher redshift is even more favoured. Hence, it seems 
unlikely that this is a lower redshift source.   

We wish to distinguish whether LEGO\_GOODS-S\#16 is an obscured AGN or a starburst 
galaxy. Several papers have presented infrared colours for obscured 
(and unobscured) AGN (Johansson et al. 2004; Lacy et al. 2004; Stern et al. 
2005; Alonso-Herrero et al. 2006) and especially two papers publish selection
criteria for obscured AGNs (Lacy et al. 2004; Stern et al. 2005). 
The colours of LEGO\_GOODS-S\#16 are inconsistent with those selection
criteria and we therefore rule out an AGN nature of this
galaxy. This conclusion is further supported by the non-detection in 
X-rays. In order to study if the SED of 
the galaxy could be fitted
by a starburst spectrum, we tried to fit the SED with a GRASIL
(Silva et al. 1998) model of a starburst galaxy. GRASIL is a
spectral stellar synthesis code, which takes into account the dust
obscuration of starlight in both molecular clouds and the diffuse medium.
Hence it is perfectly suited for the investigation of starburst
galaxies. We could fit the photometric data of LEGO\_GOODS-S\#16 by a
relatively old burst ($\sim1$ Gyr) at $z = 3.15$ with a significant amount 
of dust.
The results are shown in Fig.~\ref{grasilmodels}.
As can be seen in the Figure, these models reproduce the trends in
the observed SED relatively well. 
This appears to be a redshift $z = 3.15$ dusty starburst galaxy,
with a region where the dust amount is smaller and Ly$\alpha$ emission can 
escape, offset from the central parts of the galaxy. It would be of great 
interest to get sub-mm imaging of this object in order to constrain the SED 
better. 

\section{Comparison to Lyman-Break Galaxies}
\begin{figure}[!ht] 
\begin{center}
\epsfig{file=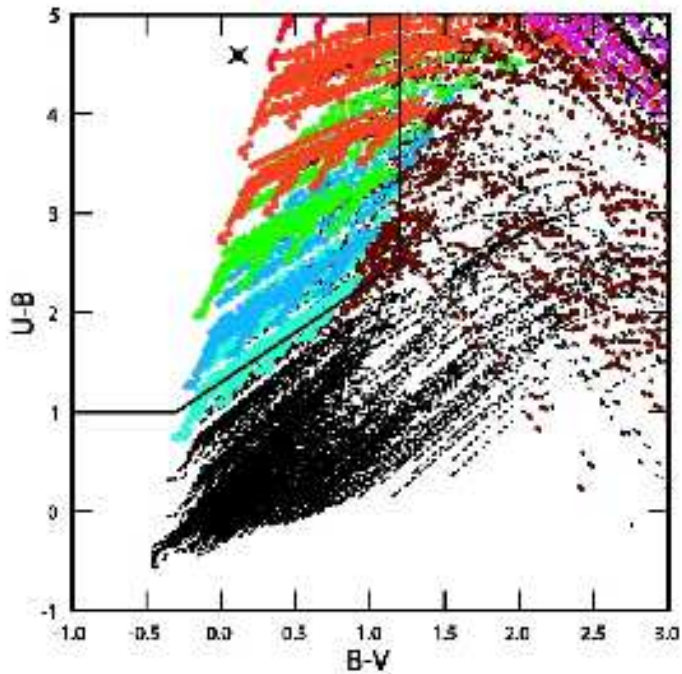,width=9.0cm,clip=,angle=90}
\caption{Colour-colour plot from Wadadekar et al. (2006) showing simulated
galaxy colours, see that paper for details. The solid line marks the area (upper 
left corner) where redshift $z \approx 3$ LBG reside. The large star in the 
upper left corner marks the colours of our best-fit synthetic model, well within
the selection boundaries for high-redshift LBGs.}
\label{lbgcol2}
\end{center} 
\end{figure}

\begin{figure*}[!ht] 
\begin{center}
\epsfig{file=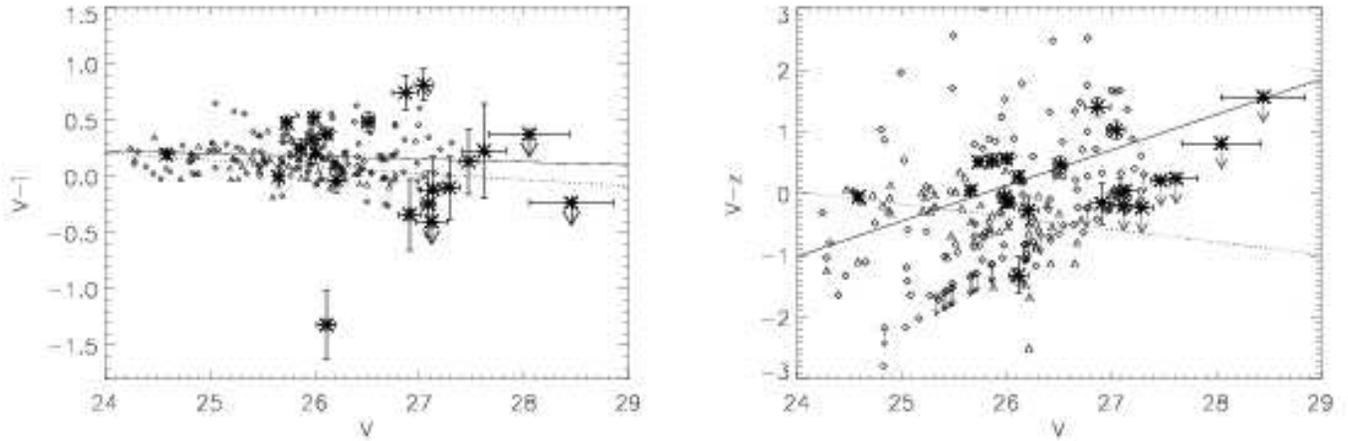,width=6.5cm,clip=,angle=90}
\caption{Colours of our candidates (stars with error bars) compared to colours 
of the sample of
faint LBGs of Wadadekar et al. (2006; diamonds) and Madau et al. 
(1996; triangles). Lines represent the best fit
to the LBG data, the solid line the fit to the sample by Wadadekar et al. (2006) 
and the dotted line the fit to the Madau et al. (1996) sample. 
\emph{Left} $V$ minus $i$ colours, \emph{Right} 
$V$ minus $z'$ colours. Small arrows indicate upper limits for the 
Wadadekar et al. (2006) sample.}
\label{lbgcol}
\end{center} 
\end{figure*}

We wish to compare our sample of LEGOs to a sample of faint Lyman Break Galaxies 
(LBGs) in order to determine the similarities and differences of the two
populations of high-redshift galaxies. First, we want to know 
if our LEGOs would be detected as LBGs and so we apply the LBG selection 
criteria for $U$-band drop-outs of Wadadekar et al. 
(2006; $U-B > 1.0$, $U-B > B-V + 1.3$ and 
$B-V < 1.2$) as well as the criteria of Madau et al. (1996; $U-B > 1.3$, 
$U-B > B-i + 1.2$ and $B-i < 1.5$) to our sample. However, our $U$-band data,
and in the case of the faintest candidates also the HST data, is
too shallow to get a useful measurement on the $U-B$ colour. Instead, we take the
best fit spectrum from the SED fitting (see Fig.~\ref{plotsed}) and convolve
it with the $U$ (F300W), $B$ and $V$ filter sensitivities and calculate the
colours. For this model spectrum, these colours become $U-B = 4.51$, 
$B-V = 0.24$ and $B-i = 0.69$ which well satisfy the selection criteria for 
$U$-band drop-outs, see Fig.~\ref{lbgcol2}. However, many of our LEGOs are 
very faint and the stacked
$B$ magnitude is fainter than the lower limit of the selection of Madau et al. 
(1996), and about half of our sample are fainter than the $V$ cut-off in the 
sample of Wadadekar et al. (2006), see Fig.~\ref{lbgcol}. 

Secondly, we wish to compare the observed optical colours (restframe UV colours)
of our LEGOs to the LBGs in order to establish if our LEGO candidates
have the same UV continuum colours as LBGs on the red side of the Lyman break.
In Fig.~\ref{lbgcol}, we plot the colours
of the two samples of faint LBGs published by Wadadekar et al. (2006) and 
Madau et al. (1996) against the colours of our candidates.
All samples are drawn from survey data-sets such as GOODS-S and HDF-N, hence
there is no bias in photometry. In the plot, we see that the LEGO candidates 
are drawn from a fainter sub-sample of the high-redshift galaxy population. 
However, for the brighter candidates among our sample, the LEGOs appear to have
UV colours similar to LBG galaxies. 

\section{Conclusion}

We have performed deep narrow-band imaging of part of the GOODS-S field.
The image revealed a set of 24 LEGO candidates, at a redshift of 
$z \approx 3.15$. Of these, three candidates have been observed 
spectroscopically and are confirmed. The spatial distribution of the candidates 
appear
to be in a filamentary structure, with a 4$\sigma$ confidence, however to 
confirm this and to plot the filament in 3D-space, we would need spectroscopic
redshifts. We have studied the entire candidate sample in all bands available 
from X-rays to infrared in the GOODS-S data-set. From the SED fitting we
conclude that the LEGOs on average have low metallicity ($Z/Z_{\odot} = 0.005$),
have stellar masses in the range of $1 - 5 \times 10^9$~M$_{\odot}$ 
and low dust extinction (A$_V \sim 0.3$).
The candidates have ages in the range of 100~--~900~Myrs . We also
find one galaxy, LEGO\_GOODS-S\#16, which is best fit by a dusty starburst 
galaxy at $z = 3.15$ 
with Ly$\alpha$-emission escaping from an area slightly offset from the central
core. 

The comparison
to a sample of $U$-band drop-out galaxies in the GOODS-S field show that the
colours of LEGOs are consistent with the selection criteria for $U$-band 
drop-outs except they are too 
faint to be detected from their continuum flux. They also have colours similar
to those of LBGs at redshift $z \approx 3$. 
In agreement with previous results (e.g. Gawiser et al. 2006), we conclude 
that Ly$\alpha$-emitters at redshift $z \sim 3.1$ are dust- and AGN-free, 
star-forming galaxies with small to medium masses. 

\begin{acknowledgements}
KN and KP gratefully acknowledges support from IDA~--~Instrumentcenter for 
Danish Astrophysics. The Dark Cosmology Centre is funded by the DNRF. The 
authors thank the DDT panel and the ESO Director General for granting time for 
follow-up spectroscopy. KN wishes to thank Matthew Hayes and Christian
Tapken for interesting 
discussions and comments on the manuscript. LFG acknowledges financial support
from the Danish Natural Sciences Research Council.
\end{acknowledgements}


\begin{thebibliography}{99}
\bibitem{alonso}Alonso-Herrero, A., P{\'e}rez-Gonz{\'a}lez, P.G., Alexander, D.M., et al., 2006, ApJ, 640, 167
\bibitem{BA1996}Bertin, E., Arnouts, S. 1996, A\&AS, 117, 393
\bibitem{Bolzonella}Bolzonella, M., Miralles, J.-M., Pell{\'o}, R., 2000, A\&A 363, 476
\bibitem{brockle}Brocklehurst, M., 1971, MNRAS, 153, 471
\bibitem{Bruzual} Bruzual G.A., Charlot S., 2003, MNRAS, 344, 1000
\bibitem{Chapman}Chapman S.C., Scott D., Windhorst R.A. et al., 2004, ApJ, 606, 85
\bibitem{Charlot}Charlot, S., \& Fall, S.M, 2000, ApJ, 539, 718
\bibitem{cowie}Cowie, L.L., \& Hu, E.M., 1998, AJ, 115, 1319 
\bibitem{Dey}Dey A., Bian C., Soifer B.T. et al., 2005, ApJ, 629, 654
\bibitem{francis2}Francis, P.J., Woodgate, B.E., Warren, S.J. et al., 1996, ApJ, 457, 490
\bibitem{Francis}Francis P.J., Williger G.M., Collins N.R. et al., 2001, ApJ, 554, 1001
\bibitem{freed}Freedman, W.L., Madore, B.F., Gibson, B.K., et al., 2001, ApJ, 553, 47
\bibitem{Fruchter06}Fruchter, A. S., Levan, A. J., Strolger, L. et al., 2006, Nature, 441, 463
\bibitem{fujita}Fujita, S.S., Ajiki, M., Shioya, Y., et al., 2003, AJ, 125, 13
\bibitem{furl}Furlanetto, S.R., Schaye, J., Springel, V., \& Hernquist, L., 2003, ApJ, 599, L1
\bibitem{FMW1999}Fynbo, J.P.U., M{\o}ller, P., \& Warren, S.J. 1999, MNRAS, 305, 849
\bibitem{FMT2001}Fynbo, J.P.U., M{\o}ller, P., \& Thomsen, B. 2001, A\&A 374, 443
\bibitem{Fynbo03}Fynbo, J.P.U., Ledoux, C., M{\o}ller, P., Thomsen, B., \& Burud, I., 2003, A\&A, 407, 147
\bibitem{Fynbo05}Fynbo, J.P.U., Gorosabel, J., Smette, A., et al., 2005, ApJ, 633, 317
\bibitem{gaw}Gawiser, E., Van Dokkum, P.G., Gronwall, C., et al., 2006, ApJL, 642, 13
\bibitem{goods}Giavalisco, M., Ferguson, H.C., Koekemoer, A.M., et al., 2004, ApJ, 600, L93
\bibitem{gilks}Gilks, W.R., Richardson, S., \& Spiegelhalter, D.J., 1995, Markov Chain Monte Carlo in Practice, Chapman \& Hall, ISBN 0412055511
\bibitem{haya}Hayashino, T., Matsuda, Y., Tamura, H., et al., 2004, AJ, 128, 2073
\bibitem{hayes}Hayes, M., \& {\"O}stlin, G., 2006, A\&A, 460, 681
\bibitem{hu}Hu, E.M., \& McMahon, R.G., 1996, Nature, 382, 231 
\bibitem{ivison00}Ivison, R.J., Smail, I., Barger, A.J., et al., 2000, MNRAS, 315, 209 
\bibitem{ivison}Ivison, R.J., Greve, T.R., Serjeant, S. et al., 2004, ApJS, 154, 124
\bibitem{jau}Jaunsen, A.O., Andersen, M.I., Hjorth, J., et al., 2003, A\&A, 402, 125
\bibitem{johansson}Johansson, P.H., V{\"a}is{\"a}nen, P., \& Vaccari, M., 2004, A\&A, 427, 795
\bibitem{Keel}Keel W.C., Cohen S.H., Windhorst R.A., Waddington I., 1999, AJ, 118, 2547
\bibitem{kenni}Kennicutt, R.C., 1983, ApJ, 272, 54
\bibitem{klaas}Klaas, U., Haas, M., M{\"u}ller, S.A.H., et al., 2001, A\&A, 379, 823
\bibitem{Lacy}Lacy, M., Storrie-Lombardi, L.J., Sajina, A., et al., 2004, ApJS, 154, 166
\bibitem{lei}Leitherer, C., Schaerer, D., Goldader, J.D., et al., 1999, ApJS, 123, 3
\bibitem{Low}Lowenthal, J. D., Hogan, C. J., Green, R. F., et al., 1991, ApJ, 377, L73
\bibitem{madau95}Madau, P., 1995, ApJ, 441, 18
\bibitem{Madau}Madau, P., Ferguson, H.C., Dickinson, M.E., et al., 1996, MNRAS, 283, 1388
\bibitem{MR2002}Malhotra, S., \& Rhoads, J.E. 2002, ApJL, 565, L71
\bibitem{Matsuda}Matsuda Y., Yamada T., Hayashino T. et al., 2004, AJ, 128, 569
\bibitem{Matsuda2}Matsuda, Y., Yamada, T., Hayashino, T. et al., 2005, ApJ, 634, L125
\bibitem{monaco}Monaco, P., M{\o}ller, P., Fynbo, J.P.U., et al., 2005, A\&A, 440, 799
\bibitem{MW1993}M{\o}ller, P., \& Warren, S.J. 1993, A\&A 270, 43
\bibitem{MW1998}M{\o}ller P., \& Warren S.J. 1998, MNRAS 299, 661
\bibitem{MF2001}M{\o}ller, P., \& Fynbo, J.U. 2001, A\&A, 372, L57
\bibitem{MW2002}M{\o}ller, P., Warren, S.J., Fall, S. M., Fynbo, J.U.,\& Jakobsen, P. 2002, ApJ, 574, 51
\bibitem{BLOB}Nilsson, K., Fynbo, J.P.U., M{\o}ller, P., Sommer-Larsen, J., \& Ledoux, C., 2006, A\&A 452, L23
\bibitem{Ouchi03}Ouchi, M., Shimasaku, K., Furusawa, H. et al., 2003, ApJ, 582, 60 
\bibitem{Ouchi04}Ouchi, M., Shimasaku, K., Okamura, S. et al., 2004, ApJ, 611, 685
\bibitem{Overz}Overzier, R.A., Miley, G.K., Bouwens, R.J., et al., 2006, ApJ, 637, 58
\bibitem{Palunas}Palunas P., Teplitz H.I., Francis P.J., Williger G.M., Woodgate B.E., 2004, ApJ, 602, 545
\bibitem{Partridge}Partridge, R.B., \& Peebles, P.J.E., 1967, ApJ, 147, 868
\bibitem{Pascarelle96}Pascarelle, S.M., Windhorst, R.A., Driver, S.P., Ostrander, E.J., \& Keel, W.C., 1996, ApJ, 456, L21
\bibitem{pea}Peacock, J.A., 1983, MNRAS, 202, 615
\bibitem{petit}Petitjean, P., Pecontal, E., Valls-Gabaud, D., \& Charlot, S., 1996, Nature, 380, 411
\bibitem{Prit}Pritchet, C.J., 1994, PASP, 106, 1052
\bibitem{Pope}Pope, A., Borys, C., Scott, D., et al., 2005, MNRAS, 358, 149
\bibitem{Schaerer}Schaerer, D., \& Pell{\'o}, R., 2005, MNRAS, 362, 1054
\bibitem{schmitt}Schmitt, H.R., Calzetti, D., Armus, L., et al., 2006, ApJ, 643, 173
\bibitem{silva}Silva, L., Granato, G.L., Bressan, A., \& Danese, L., 1998, ApJ, 509, 103
\bibitem{smail}Smail, I., Chapman, S.C., Blain, A.W., \& Ivison, R.J., 2004, ApJ, 616, 71
\bibitem{Steidel}Steidel, C.C., Adelberger, K.L., Giavalisco, M., Dickinson, M., \& Pettini, M., 1999, ApJ, 519, 1
\bibitem{SAS2000}Steidel, C.C., Adelberger, K., Shapley, A.E., et al.,
2000, ApJ, 532, 170
\bibitem{Stern}Stern, D., Eisenhardt, P., Gorjian, V., et al., 2005, ApJ, 631, 163
\bibitem{Venemans}Venemans, B.P., R{\"o}ttgering, H.J.A., Miley, G.K. et al., 2005, A\&A, 431, 793
\bibitem{Wadadekar}Wadadekar, Y., Casertano, S., \& de Mello, D., 2006, AJ, 132, 1023
\bibitem{Wang}Wang, J.X., Rhoads, J.E., Malhotra, S., et al., 2004, ApJL, 608, 21 
\bibitem{Warren96}Warren, S.J., \& M{\o}ller, P., 1996, A\&A, 311, 25
\bibitem{Warren01}Warren, S.J., M{\o}ller, P., Fall, S.M.,\& Jakobsen, P., 2001, MNRAS, 326, 759
\bibitem{W2002}Weidinger, M., M\o ller, P., Fynbo, J.P.U., Thomsen,
B., \& Egholm, M.P. 2002, A\&A, 391, 13
\bibitem{wolfe86}Wolfe, A.M., Turnshek, D.A., Smith, H.E, \& Cohen, R.D., 1986, ApJS, 61, 249
\bibitem{wolfe05}Wolfe, A.M., Gawiser, E., \& Prochaska, J.X., 2005, ARAA, 43, 861
\end{thebibliography}
\end{document}